\documentclass[%
reprint,
superscriptaddress,
 amsmath,amssymb,
 aps,
floatfix,
showkeys
]{revtex4-1}

\usepackage[usenames, dvipsnames]{color}
\usepackage{graphicx}% Include figure files
\usepackage{dcolumn}% Align table columns on decimal point
\usepackage{bm}% bold math
\usepackage{url}
\usepackage{times}
\usepackage{textcomp}
\usepackage{multirow}
\begin{document}

\title{Machine-learning models for Raman spectra analysis of twisted bilayer graphene}% Force line breaks with \\
\author{Natalya Sheremetyeva}
\affiliation{%
 Department of Physics, Applied Physics, and Astronomy, Rensselaer Polytechnic Institute, Troy, USA
}
% \email{sheren@rpi.edu}
%%
\author{Michael Lamparski}%
\affiliation{Department of Physics, Applied Physics, and Astronomy, Rensselaer Polytechnic Institute, Troy, USA
}

\author{Colin Daniels}%
\affiliation{Department of Physics, Applied Physics, and Astronomy, Rensselaer Polytechnic Institute, Troy, USA
}

\author{Benoit Van Troeye}
\affiliation{%
 Department of Physics, Applied Physics, and Astronomy, Rensselaer Polytechnic Institute, Troy, USA
}

\author{Vincent Meunier}
\affiliation{%
Department of Physics, Applied Physics, and Astronomy, Rensselaer Polytechnic Institute, Troy, USA
}%
\email{meuniv@rpi.edu}

% \date{\today}% It is always \today, today,
%              %  but any date may be explicitly specified

\begin{abstract}
The vibrational properties of twisted bilayer graphene (tBLG) show complex features, due to the intricate energy landscape of its low-symmetry configurations. A machine learning-based approach is developed to provide a continuous model between the twist angle and the simulated Raman spectra of tBLGs. Extracting the structural information of the twist angle from Raman spectra corresponds to solving a complicated inverse problem. Once trained, the machine learning regressors (MLRs) quickly provide predictions without human bias and with an average 98\% of the data variance being explained by the model. The significant spectral features learned by MLRs are analyzed revealing the intensity profile near the calculated G-band to be the most important feature. The trained models are tested on noise-containing test data demonstrating their robustness. The transferability of the present models to experimental Raman spectra is discussed in the context of validation of the level of theory used for construction of the analyzed database. This work serves as a proof of concept that machine-learning analysis is a potentially powerful tool for interpretation of Raman spectra of tBLG and other 2D materials. 
\end{abstract}
                             % Classification Scheme.
 %Use showkeys class option if keyword
                              %display desired
\maketitle

\section{\label{sec:intro}Introduction}
Graphene, as the archetypal 2D material, has been studied extensively in the last two decades and new physical phenomena are being discovered to this day~\cite{Bianco2020}. When graphene layers are stacked on top of each other, interlayer interactions lead to distinct dimensional crossovers from 2D to 3D via Bernal stacked (AB) bilayer to multilayer, and finally graphite. Moreover, defects and disorders can drastically modify the interlayer interactions resulting in vastly distinct material properties~\cite{Mogera2020}. One such disorder can be produced by an angular twist between layers such that the layers stack away from the conventional AB stacking order~\cite{Mogera2020}. The resulting twisted bilayer graphene (tBLG) has been shown to exhibit exciting new properties, as \textit{e.g.} unconventional superconductivity and ferromagnetism for a small twist angle of about $\theta=1.1^{\circ}$, among other $\theta$-dependent phenomena~\cite{Cao2018,Cao2018b,Yankowitz2019,Sharpe2019,Codecidoeaaw2019}. Such twist angle dependent properties of tBLGs make these systems attractive for potential applications. However, before exploitable in application, tBLG must be precisely characterized and control of the twist angle is necessary.

In general, tBLG can exist with any arbitrary twist angle, either in an incommensurate or commensurate form. Here, we investigate the commensurate subset~\cite{shallcross2010,Lamparski2020}.
Experimentally, different synthesis techniques have been developed to produce tBLGs with a given $\theta$, see Ref.~\cite{Mogera2020} for an extensive review. However, combining exact angle control and the synthesis of large-area, small twist angle tBLG with uniform stacking orientation over the whole domain-range remains a challenge~\cite{Chen2016,Wang2017}. 

Transmission electron microscopy (TEM) and electron diffraction are typically used for characterization of the experimentally synthesized tBLG structures with respect to their twist angle~\cite{Kim2012,Jorio2013,Chen2016,Wang2017}. Experimental Raman spectra of tBLG have been analyzed as well, revealing complicated, non-monotonic and resonant fingerprints of the positions, intensities, and widths (Full Width at Half Maximum (FWHM)) of the prominent G- and 2D- bands~\cite{Kim2012}. Additionally, new families of Raman bands have been reported in tBLG with shifts in a large range below and around the G-band. These bands are denoted as R bands and have been attributed to phonons in the interior of the Brillouin Zone (BZ) of single layer graphene (SLG) being activated by a $\theta$-dependent $\mathbf{q}$-vector in tBLG which makes them accessible by first-order Raman scattering~\cite{Jorio2013}. The R bands include the layer-breathing mode vibrations and their positions are highly sensitive to the twist angles. They also exhibit intricate resonance effects. For this reason, these bands have been suggested and used as signatures of tBLG \cite{Jorio2013,Chen2016}. 

Some theoretical efforts in explaining the subtle twist angle dependent phenomena in tBLG's Raman spectra have been undertaken \cite{Coh2013,Popov2018}. Coh et al. have focused on the resonant intensity enhancements of the G and 2D peaks using a tight-binding approach and procedures to deduce $\theta$ from a set of Raman spectra recorded with multiple laser lines for the same sample have been suggested \cite{Coh2013}. More recently, Popov used a non-orthogonal tight-binding model to address the resonant Raman signature of the G-band. An analytical model of the G-band's intensity as a function of $\theta$ was then deduced via fitting of the obtained intensities. However, this model turned out not to be universally applicable for an arbitrary excitation laser energy \cite{Popov2018}. Both works provided valuable accounts of theoretical simulations of tBLG's Raman spectra and were able to reproduce experiments with a high degree of agreement. However, the number of studied twisted structures was limited to a few tens and structures with very small $\theta$-angles ($<5^{\circ}$) could not be included due to prohibitive computational cost. Moreover, the G-band's position was assumed to be constant. While the experimental G-band's position does indeed not shift as considerably as that of the 2D-band with $\theta$, it does exhibit small non-monotonic variations of a couple cm$^{-1}$ (see \textit{e.g.} Fig. 2a in Ref. \onlinecite{Kim2012}). With improving Raman spectroscopy hardware such small differences can increasingly be resolved and might prove significant \footnote{Recent theoretical and experimental studies suggest that a significant geometrical relaxation takes place in small-$\theta$ tBLGs~\cite{Alden2013,Nam2017,Wijk2015,Dai2016,Jain2016,Choi2018,Koshino2019b,Yong2019,Gargiulo2017,Angeli2018,Guinea2019,Lucignano2019}, which should in principle translate into a shift of the G-band.}.

To the best of our knowledge, a model establishing a one-to-one correspondence between one given Raman spectrum, including the peak position variations, and the underlying twist angle of tBLG is missing to date. A direct mapping between Raman spectra and $\theta$ is of significant practical relevance as Raman spectroscopy is a fast and non-destructive method rapidly gaining popularity for characterization of other 2D materials as well \cite{Liang2017}.

Abstractly, a mapping between a Raman spectrum ($\mathbf{y}$) and the atomic arrangement ($\mathbf{x}$) in a tBLG sample is given by some physical (quantum-mechanical) laws ($f$) as $\mathbf{y}=f(\mathbf{x})$. The twist angle $\theta$, as a proxy to structural information \cite{Lamparski2020}, should thus in principle be accessible via inversion of the problem: $\theta=f^{-1}(\mathbf{y})$. As discussed above, the experimental Raman spectral information is highly nontrivial, explaining why the "traditional" inverse-problem solution is based on establishing semi-quantitative fingerprints.

In contrast, machine learning (ML) approaches are well adapted to help finding a quantitative solution to the above inverse problem of tBLG characterization using Raman spectroscopy. ML methods have experienced a surge of popularity in the fields of condensed-matter physics, chemistry, and materials science recently. They have been applied successfully to a plethora of problems, such as solutions of many-body problems \cite{Carleo2017}, bypassing density-functional calculations \cite{Brockherde2017}, force-field potentials generation \cite{Behler2007}, and evaluation of experimental atomic force microscopy (AFM) images \cite{Borodinov2020}. The impressive and growing track-record of ML approaches can be linked to their data-driven nature and the ability to provide quantitative predictions of complicated correlations that are free of human bias. Another inverse characterization problem, similar to that of interest in this work, has been addressed by Carbone et al. in the context of computational x-ray absorption near-edge structure (XANES) spectra. There, useful links to local chemical environments were established via supervised ML classification \cite{Carbone2019}.
 
In this work, we addressed the inverse-problem of linking calculated Raman spectra to the twist angle of tBLG using supervised ML regression. Specifically, non-linear regression was used to establish a continuous model of $\theta(\text{spectrum})$ using the computational Raman spectra of a large number of tBLG structures as input features. The importance of spectral features for the training of the models was systematically explored using dimensionality reduction and an in-depth analysis of the decision-tree based regression. The ML models trained here on synthetic Raman spectra serve as a proof of concept for such an analysis being promising for spectral data of similar general shape. Thus, the present work is a first step on the path to a potentially effective characterization tool using Raman spectra of tBLGs.  

\section{\label{sec:methods}Methods}

\subsection{\label{subsec:Amethods}Data Acquisition}
The construction of the database for the present analysis consists of three main steps. First, a large number of tBLG superlattices are constructed and their atomic positions are carefully relaxed. Second, the phonon eigenfrequencies and eigenvectors of the relaxed structures are calculated via diagonalization of the corresponding dynamical matrices. Lastly, Raman intensities of the calculated phonon modes are obtained using a semi-empirical bond-polarizability model \cite{guhaEmpiricalBondPolarizability1996,saitoRamanSpectraGraphene2010}. 

The procedure used for construction and relaxation of the tBLG superlattices is detailed in Ref. \cite{Lamparski2020}. Due to symmetry, nonequivalent twist angles are limited to the range $0^{\circ}\le\theta<30^{\circ}$. Restricting the supercells to contain fewer than 20,000 atoms results in the construction of 692 commensurate tBLG structures. 

Classical force-fields are used for structural relaxation and phonon calculations in the harmonic approximation. The force-fields are built as a sum of the intra- and interlayer forces. The former are modeled using the second-generation  REBO  potential \cite{Brenner2002},  while the latter are computed using  the  registry-dependent  Kolmogorov–Crespi  (KC)  potential \cite{Kolmogorov2005},  in  its  local  normal  formulation \cite{Lamparski2020}.

Due to intricacies of structural relaxation of tBLG, some phonon modes of some structures in the database have complex eigenfrequencies. This happens particularly for structures with a small twist angle  (\textit{i.e.}, large supercells) where perfect relaxation was not computationally feasible due to extremely shallow local minima in the energy landscape~\cite{Lamparski2020}. However, the remaining imaginary frequencies have small magnitudes, $|\omega|<4$ cm$^{-1}$, that were assumed to be negligible. In addition, from the real phonon modes, only frequencies $>1$ cm$^{-1}$ were included during the construction of the Raman spectra in the data preprocessing step (see next section).

Finally, non-resonant, first-order Raman intensities are calculated in the Plazcek approximation \cite{Placzek1959,Umari2001,Ceriotti2006,Liang2014} combined with the semi-empirical bond polarizability model~\cite{cardonaLightScatteringSolids1982,guhaEmpiricalBondPolarizability1996}. The latter was used to calculate the change in polarizability of tBLG induced by its vibrational modes. The bond polarizability model is computationally inexpensive and has been used successfully in modeling of low-frequency modes in stacked 2D materials~\cite{liangInterlayerBondPolarizability2017}, as well as in the analysis of overall trends in Raman spectra of finite-size graphene nanoribbons~\cite{OverbeckUniversalLengthDependentVibrational2019}. The specific model parameters used here are given in table S2 in Supplemental Material (SM) [link to SM inserted by publisher]. A typical experimental laser back-scattering set-up is assumed throughout this study and the calculated Raman intensities are averaged over the in-plane laser polarization angles.

\subsection{\label{subsec:Bmethods}Spectral Data Preprocessing}

Using the calculated phonon frequencies and their associated Raman intensities, Raman spectra are constructed as a sum of Lorentzian curves sampled on a wavenumber grid with step size of 0.25 cm$^{-1}$. As stated previously, frequencies below 1 cm$^{-1}$ are excluded. Each Lorentzian is centered on a calculated phonon mode with the amplitude given by the calculated Raman intensity. Raw intensities smaller than 1\% of the maximum intensity in a given spectrum were set to zero in order to speed up the spectrum construction. This turned out to be especially important for systems with numerous phonon modes (supercells with many atoms). The same FWHM of 0.5 cm$^{-1}$ is used for all Lorentzian peaks. The full Raman spectrum of each system is normalized to the strongest feature and is thus assigned an arbitrary intensity of 1. Thus, the relative intensities between different peaks in a given spectrum is preserved. Note that this normalization procedure results in the loss of information about relative intensities between spectra of systems with different twist angles. It should also be noted here, that the relative intensities within a spectrum given by the semi-empirical bond polarizability model might not be very reliable for reproducing absolute experimental intensities (a more in-depth discussion of the theory validation is given in Sec.~\ref{subsec:Eresults}). However, the purpose of this approach is the identification of Raman active phonon modes rather than a full experiment replication.

\subsection{\label{subsec:Cmethods}Training Machine Learning Models}
Different ML regression algorithms, as implemented in the python library \textsc{scikit-learn} \cite{scikit-learn2011}, were tested and their performance was compared. Key differences between these algorithms will be outlined in Sec.~\ref{subsec:Cresults} when discussing their performance in more detail. Each algorithm's internal parameters were optimized for best performance using a grid-search with ten-fold cross validation (CV) on the training set. Cross-validation is a paramount part of any ML analysis as it provides insights into how the trained model will perform on data it has not seen in training. In the $k$-fold CV approach, the complete data set is (evenly) split into $k$ groups (folds) one of which is used as the held-out set for validation and the remaining ones are used for fitting \cite{James2013}. Repeating the training and validation cycle $k$ times and averaging over the resulting performance scores provides a statistically robust measure of the model's performance. Optimizing the internal parameters of an ML model with the incorporation of the CV step allows for minimizing the potential for the model to overfit on the training set \footnote{Overfitting occurs when the fitting coefficients are highly optimized to capture the training data, but fail to adequately represent previously unseen data.}. Here, the total data set was split into a training and a test sets at random with 600 and 92 samples, respectively. This corresponds to a relative splitting of 87/13 (compared to typical splittings of 80/20 or 90/10). In this way, the 600 training samples could be evenly divided into subsets for cross-validation. Most of the algorithms compared here use the mean-squared-error (mse) as a criterion for model optimization (i.e., loss-function). For performance comparison the average CV and test $R^{2}$ coefficients, average CV mean-squared error (MAE), and root-mean-squared error (RMSE) were used as metrics. 

%%%%%%%%%%%%%%%%%%%%%%%%%%%%%%
\section{\label{sec:results}Results and Discussion}
The present analysis consists of the following key steps: Visual inspection of the generated spectral database, principal component analysis (PCA) of the database, training of different machine-learning regressors (MLRs) and comparison of their performance, and the discussion of the importance of features for the solution of the posed inverse problem. Finally, we discuss challenges of transferring the models for analysis of experimental Raman spectra. 

\subsection{\label{subsec:Aresults}Visual Inspection of the Spectral Database}
%%%--------------------
\begin{figure*}[ht!]
\includegraphics[width=1.\textwidth,keepaspectratio]
{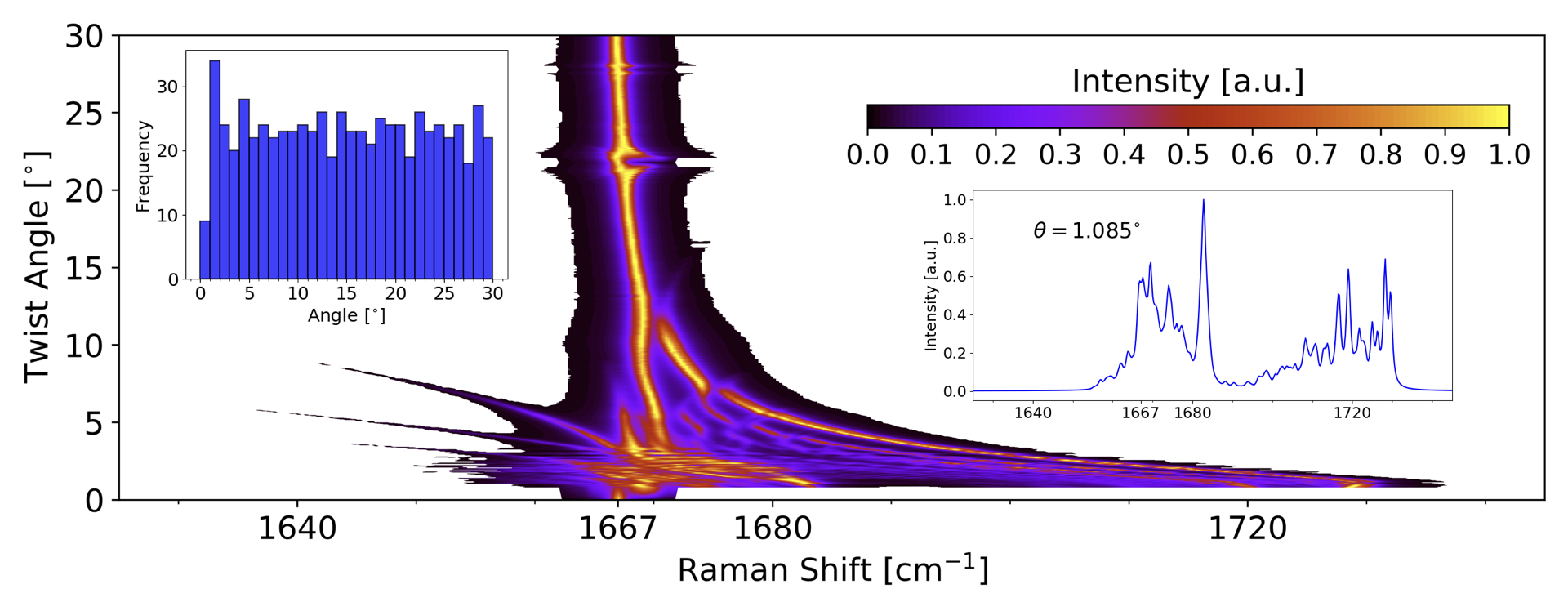}
\caption{\label{fig:Spectra}
Calculated Raman spectra for different twist angles $\theta$ constituting the spectral database. Only the HF range around the calculated G-band of untwisted BLG (calculated peak location at 1667 cm$^{-1}$) is shown. Full spectra are given in SM [link to SM], where most of the remaining frequency range contains no noticeable Raman intensities for most systems. Zero intensity is indicated by the white color. The FWHM of the Lorentzian spectral peaks is 0.5 cm$^{-1}$. The figure-inset on the right shows an example spectrum for $\theta=1.085^{\circ}$ showing multiple additional peaks not present for $\theta=0^{\circ}$. The figure-inset on the left shows twist angles distribution in the structural database with 1$^{\circ}$ histogram bins.
}
\end{figure*}
%%%--------------------

Figure \ref{fig:Spectra} shows the calculated Raman spectra for different twist angles $\theta$. The wave-number region is limited to the high-frequency (HF) range in the vicinity of the calculated G-band of untwisted BLG. This is where most of the intensity information is contained. Full spectra can be viewed in SM [link inserted by publisher]. Figure S1 shows that most Raman spectra have practically zero intensity below 1625 cm$^{-1}$. Only structures with $\theta<5^{\circ}$ show non-zero Raman intensities in the low-frequency (LF) range below 20 cm$^{-1}$. As will be shown in the next section, removing the spectral information below 1625 cm$^{-1}$ does not alter the results of the ML based analysis.

Visual inspection of figure \ref{fig:Spectra} allows for detection of some trends with twist angle. The calculated G-band of Bernal bilayer graphene is located at 1667 cm$^{-1}$ (experimental location at about 1580 cm$^{-1}$~\cite{Mogera2020}) and is the only Raman peak in the spectrum for $\theta=0^{\circ}$.
For small twist angles, many additional peaks emerge compared to Bernal bilayer graphene. This is due to the corresponding structures formally being described by large supercells with thousands of atoms that support large numbers of vibrational modes, some of these modes becoming Raman active due to symmetry breaking. Because of the large number of Raman active modes, the definition of the G-band becomes ambiguous in those systems. For simplicity, for every $\theta$ we define the corresponding G-band as the most intense peak in the corresponding spectrum \footnote{For a rigorous G-band vs. $\theta$ identification a more involved phonon unfolding would be required.}. For twist angles below 5$^{\circ}$, the G-band positions are blue shifted compared to the $\theta=0^{\circ}$ configuration. For some twist angles in that range, the shift magnitude is erratic, while for others the G-band's frequency $\omega(\theta)$ follows an exponentially decreasing trend (this is difficult to see in Fig. \ref{fig:Spectra}, but becomes apparent in Fig. \ref{fig:Gbandplot}).
For slightly larger angles, between 5$^{\circ}$ and 10$^{\circ}$, there are new branches of peaks both above and below the G-band's feature. The locations of the peaks in these branches mostly show frequency red-shift trends with $\theta$. Above 10$^{\circ}$ the most intense peak shifts abruptly to about 1670 cm$^{-1}$, slightly above the frequency associated with the Bernal configuration of bilayer graphene, and exhibits a piece-wise blue-shift trend up to about 13$^{\circ}$. The remaining $\theta$ region is characterized by a mostly red-shifting G-band position interrupted by intensity-discontinuities and additional emerging intensity-branch structures around 22$^{\circ}$ and 28$^{\circ}$.

Figure~\ref{fig:Spectra} indicates a quite complicated evolution of Raman spectra with twist angle. Linking a given spectrum quantitatively to $\theta$ using standard fingerprint approaches is challenging. In contrast, quantitative insights can be gained using an ML based approach as we will show in the rest of this paper. 

\subsection{\label{subsec:Bresults}Principal Component Analysis of the
Spectral Database}
%%%--------------------
\begin{figure}[ht!]
\includegraphics[width=0.5\textwidth,keepaspectratio]
{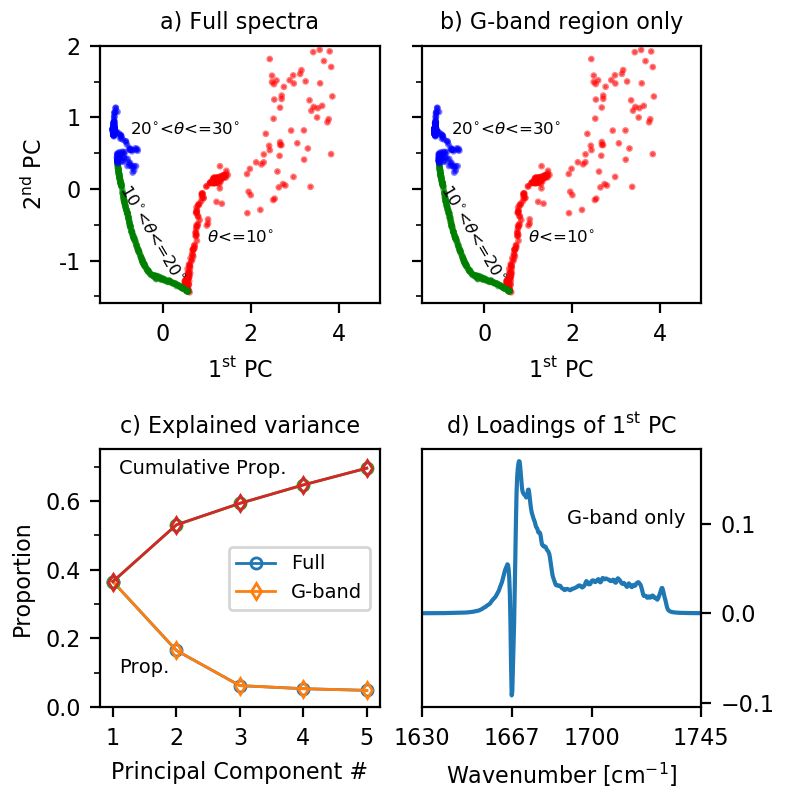}
\caption{\label{fig:PCA}
Main results of PCA analysis of the Raman spectra of tBLGs. Top row: transformed Raman spectra in the space of the first two principal components. Data set containing the full spectra a) and the spectra restricted to the G-band region only b). Different colors (online) denote the three coarse angle groups: red dots mark $\theta<=10^{\circ}$, green dots denote $10^{\circ}<\theta<=20^{\circ}$, and blue dots correspond to $20^{\circ}<\theta<=30^{\circ}$. Bottom row: key metrics of PCA. c) Proportion of explained variance for each PC and the cumulative proportion of explained variance as a function of PC number for the full spectra (open circles) and for the G-band limited spectra (open diamonds). d) Loadings $\phi_{i1}$ as a function of feature position in the frequency domain for the G-band only data set.
}
\end{figure}
%%%--------------------

PCA is an unsupervised learning technique that does not have the goal of predicting a response. Instead, it provides a useful tool for data visualization via dimensionality reduction, \textit{i.e.}, construction of a smaller number of representative features out of the complete input-features set. The new representative features are designed to explain most of the variability of the input features \cite{James2013}. In more mathematical terms, given a set of $p$ input feature vectors $X_1, X_2 \dots X_p$ for $n$ samples ($X_i$ has length $n$), the first principal component (PC) $Z_1$ is given by
\begin{align}
    Z_{1}=\phi_{11} X_{1}+\phi_{21} X_{2}+\ldots+\phi_{p 1} X_{p},
\end{align}
with the coefficients $\phi_{i1}$ being the so-called \textit{loadings} of the first PC. These loadings are elements of the first PC loading vector $\phi_1$, normalized as $\sum_{i=1}^{p} \phi_{i 1}^{2}=1$. The coefficients $\phi_{i1}$ are obtained through optimization such that the variance of all possible linear combinations 
\begin{align}
    z_{j 1}=\phi_{11} x_{j 1}+\phi_{21} x_{j 2}+\ldots+\phi_{p 1} x_{j p} \ , \ j=1 \dots n
\end{align}
is maximized across the samples. In geometrical terms, $\phi_1$ defines the direction in feature space along which the data vary the most. The second and all the following PCs are constructed subsequently in an analogous fashion with the constrain that they are orthogonal (\textit{i.e.}, uncorrelated) to the previous components \cite{James2013}. 

Here, the features X$_i$ correspond to vectors containing the spectral intensities at a given wavenumber for each twist angle $\theta$ (represented along the vertical lines through Fig.~\ref{fig:Spectra}). Note, that PCA results are sensitive to whether the individual features have been distinctly scaled as \textit{e.g.}, in case of different units. For this reason, one typically has to scale each variable to have a standard deviation of one before performing PCA \cite{James2013,Carbone2019}. However, when all variables are measured using the same unit a scaling is not necessary. Thus, a scaling was not applied here as all the intensities within a spectrum have the same units.

Figure~\ref{fig:PCA} shows the key pieces of information obtained from PCA. The top row depicts the spectra as represented by the first two PCs. Subplots \ref{fig:PCA}a and \ref{fig:PCA}b compare two different cases. In subplot a), the complete spectra in the range [1,1735] cm$^{-1}$ were used. In subplot b), the spectral range was restricted to the region surrounding the G-band in pristine BLG, [1625,1735] cm$^{-1}$. This allows to assess the importance of the LF-region features by comparing the resulting clustering patterns between the two cases. If the features near the G-band dominate the spectra, similar patterns can be expected in both subplots. 

 The points in subplots a) and b) are color-coded by three groups: $\theta<=10^{\circ}$ (red), $10^{\circ}<\theta<=20^{\circ}$ (green), and $20^{\circ}<\theta<=30^{\circ}$ (blue). Overall, there is a sizable level of clustering by color. The three resulting clusters are well separated, although there is some overlap between the groups of angles. Comparing the full (Fig.~\ref{fig:PCA}a) and the G-band (Fig.~\ref{fig:PCA}b) spectra, we note small but noticeable differences in the shapes of the three main clusters. However, on the large scale the clustering patters are almost identical. This indicates that there is essentially no information loss when reducing the full spectra to the G-band region only. This was already intuitively established in the previous section based on visual inspection, but PCA now provides a more robust and analytical support for this finding. The comparison for the two spectral data sets also provides an elemental feature-importance assessment.

An additional metric of the PCA is the proportion of explained variance for each PC and the cumulative proportion of explained variance as a function of the number of PCs (see Fig.~\ref{fig:PCA}c). Again, for both data sets the explained variance is essentially identical. The first and the second components each explain somewhat less than 40\% and 20\% of the data variance, respectively. The first five PCs together account for about 70\% of the overall variance. This indicates that most of the data variance can be explained by the first two PCs. Still, the inverse problem can be expected to be more involved as the data variance is not fully explained by a few linear combinations of spectral intensities.

Finally, Fig.~\ref{fig:PCA}d shows the coefficients $\phi_{i1}$ for the spectral data set containing only the G-band region. The largest coefficients are assigned to features close to the G-band, meaning that these are most important for explaining the variability in the spectral data set. This trend will become apparent again later in Sec.~\ref{subsec:Dresults} when we discuss the feature importance for ML regression in more detail.

\subsection{\label{subsec:Cresults}Performance of MLRs}
\begin{table}[ht!]
\centering
\begin{ruledtabular}
\begin{tabular}{ccccc}
Model & R$^2$ & MAE [$^{\circ}$] & RMSE [$^{\circ}$] & Test R$^2$ \\
KRR (lin) & 0.83 $\pm$ 0.03 & 2.39 $\pm$ 0.19 & 3.45 $\pm$ 0.45 & 0.91 \\
KRR (rbf) & 0.98 $\pm$ 0.04 & 0.26 $\pm$ 0.12 & 0.70 $\pm$ 0.92 & 0.99 \\
RFR & 0.98 $\pm$ 0.04 & 0.19 $\pm$ 0.14 & 0.72 $\pm$ 0.95 & 0.99 \\
MLP & 0.98 $\pm$ 0.04 & 0.27 $\pm$ 0.12 & 0.69 $\pm$ 0.92 & 0.99 \\
\end{tabular}
\end{ruledtabular}
\caption{\label{tab:scores}Performance metrics for different MLRs. The first three metrics $R^2$, MAE, and RMSE correspond to average values obtained in ten-fold cross-validation with the uncertainty given by the standard deviation (std). Test $R^{2}$ is reported for the hold-out set, that has not been seen by the model during training. Details regarding free-parameter settings of each ML model are given in SM.
}
\end{table}

Table \ref{tab:scores} shows different performance scores for a selected number of tested machine learning regressors used for modeling the twist angle of tBLG given its Raman spectrum. The values in the first three columns are averages over ten cross-validation cycles on the training set. The last column corresponds to the score of the trained model applied to the hold-out test set. The compared models are of linear and non-linear characters and are based on different algorithms. 

Kernel ridge regression (KRR) corresponds to the combination of ridge regression with the kernel trick \cite{scikit-learn2011,Rupp2015}. Ridge regression is a form of linear regression with additional regularization of the fitting parameters \footnote{Specifically, $l_2$ regularization: a penalty term of the form $\lambda\|\boldsymbol{\beta}\|^{2}$ is added to the ordinary least-squares fitting expression, where $\boldsymbol{\beta}$ is the vector of the regression coefficients and $\lambda$ is a hyperparameter determining the regularization strength.}. The kernel trick constitutes a mapping of the original features into a feature space induced by a kernel function \cite{scikit-learn2011,Rupp2015}. For a linear kernel this simply corresponds to linear ridge regression. When a Gaussian kernel (radial basis function - rbf) is employed, non-linear fitting can be performed. Next, the Random Forest Regressor (RFR) is an algorithm from the family of the so-called ensemble methods and is based on randomized decision trees. A single decision tree can be constructed for the purpose of regression by dividing the feature space in high-dimensional ``boxes" and taking the mean of the target in that box as the predicted value. The number and the boundaries of the boxes have to be optimized such that the loss-function (mse in this case) is minimized. A decision-tree fit results in a non-linear model. However, a single decision tree can suffer from overfitting on the training set~\cite{James2013}. To avoid overfitting, a forest of distinct trees is constructed and the average over the forest is used as the prediction. The randomized aspect of RFR comes into play when a subset of randomly selected features is used at each node (branch splitting point) for ``box"-construction. In this way, instead of having multiple potentially very similar trees in the forest, the structure of the trees is diversified leading to a more stable model \cite{scikit-learn2011,James2013}. Lastly, we used the Multi-Layer-Perceptron (MLP). The MLP belongs to a class of feedforward artificial neural networks (ANNs). It is composed of multiple layers of perceptrons (neurons) and employs an activation function that maps the weighted inputs to the output of each neuron \cite{scikit-learn2011}. The number of layers and the number of neurons in each layer are free parameters that have to be tuned for a given problem. MLPs are the most basic type of ANNs. In general, ANNs have become increasingly popular as powerful tools for problems with extremely large data sets and/or extremely many features, such as image recognition for example \cite{scikit-learn2011}. In the field of material science, ANNs have not been used as vastly due to the relatively small amount of data available to date. In addition, the highly non-linear nature of ANNs makes the interpretation of the resulting models rather impractical for gaining physical insights. Here, the MLP was used for completeness of the comparison between models.

All non-linear models perform very well with an average $R^2$ coefficient of 0.98. The accuracy of each model is at least 2$^{\circ}$ or better. In contrast, the linear model resulted in weaker performance compared to non-linear models. Thus, the relationship between input Raman spectra and target twist angles is better described by a non-linear function. Furthermore, all the non-linear models performed very well on the hold-out test set with a $R^2$ score of 0.99. However, the high-test score is not an indicator of how well the models would perform on other test sets, and is partly accidental and partly explained by the fact that our data set is overall very homogeneous. 

To further analyze the models reported in table \ref{tab:scores}, we inspected the learning curves for each model. Figure S2 shows the $R^2$ score for each model as a function of the number of samples in the training set where the training (red diamonds) and CV scores (green circles) are compared. Such learning-curves typically exhibit a number of key features. First, the training score is naturally higher than the CV score. Second, for small sample sizes the model is overfitting significantly resulting in a very high training score contrasted by a comparatively low CV score. As the training data set grows, the training and CV scores approach each other mostly due to the increasing CV score. In the ideal situation of the trained model being accurate enough on the training set, but flexible enough on the "test" set in CV procedure (\textit{Bias–variance tradeoff}~\cite{James2013}), the two scores become almost identical. All the non-linear models used here show a similar trend of increasing CV score with increasing number of training samples. However, for the maximum training size, the three models all have distinct training -- CV score differences. RFR has the largest $R_{\mathrm{train}}^2-R_{\mathrm{CV}}^2$, followed by KRR and MLP with the smallest score difference. This analysis suggests that RFR tends to slightly overfit on the training set, whereas KRR and especially MLP are less susceptible to this problem. Therefore, MLP should be applied to unseen test data. However, the cost of MLP is the low interpretability. For this reason, in order to get a more "under-the-hood" view of the ML models, the importance of features used for regression will be discussed using the RFR model as the showcase in the next section.

\subsection{\label{subsec:Dresults}Feature Importance}
%%%--------------------
\begin{figure*}[ht!]
\includegraphics[width=1.\textwidth,keepaspectratio]
{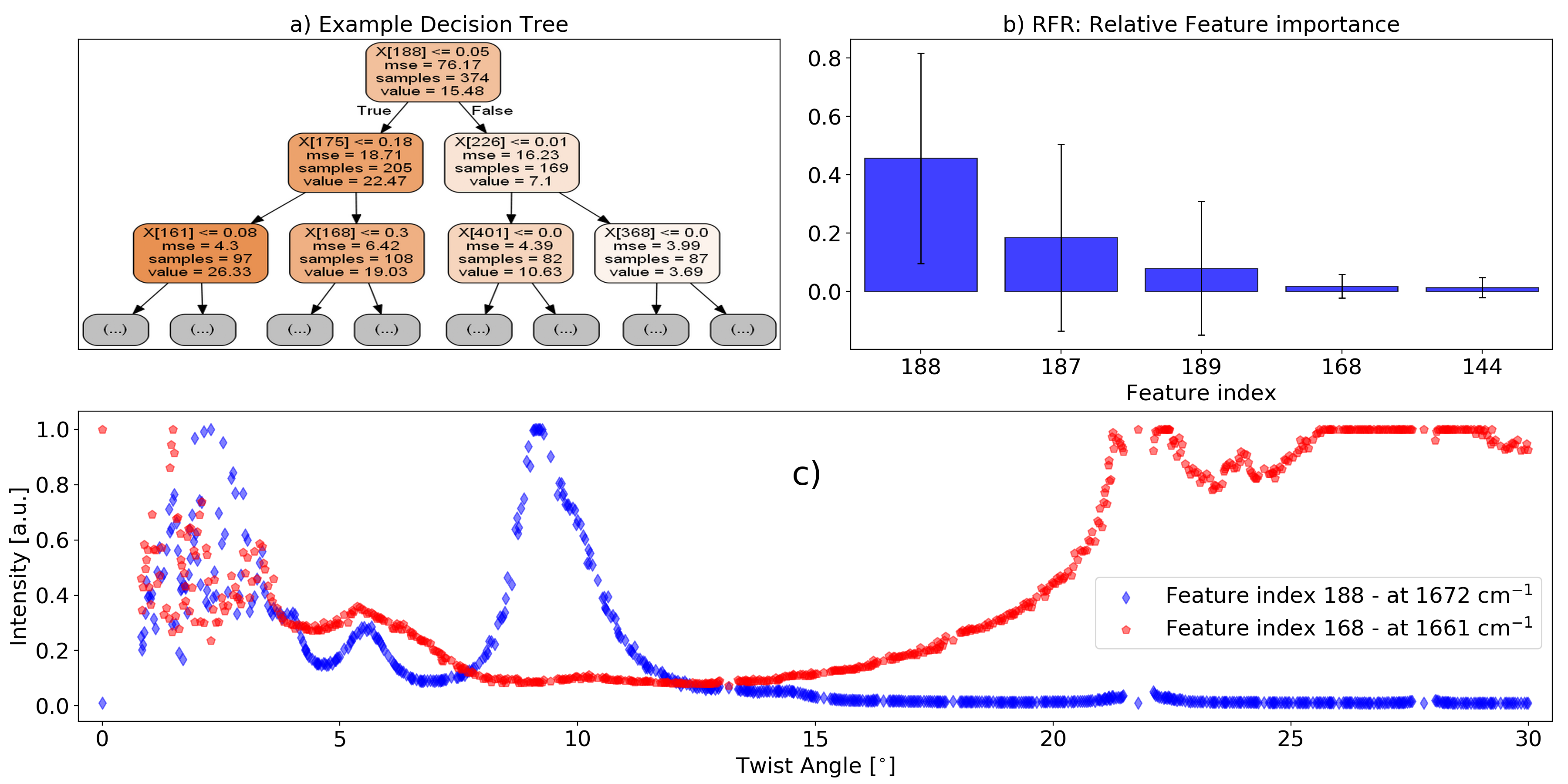}
\caption{\label{fig:RFR_features}
Feature importance analysis based on the RFR model. a) Example decision tree from the random forest. Only the first three branches are shown. b) Relative feature importance for the first five most important features across the forest. Error bars are given by the std. c) Two of the five most important features as a function of twist angle.}
\end{figure*}
%%%--------------------

Even though RFR might be prone to overfitting, its relatively simple "mechanics" allow for a view inside the model to understand what features are most important for learning the twist angle from Raman spectra. Figure \ref{fig:RFR_features} shows key information regarding the RFR model. Subplot \ref{fig:RFR_features}a shows an example decision tree with its top three nodes. The feature at the top of the tree corresponds to the most important one for separating the data set. Subplot \ref{fig:RFR_features}b shows the relative feature importance for the five most important features evaluated across the full forest. Finally, subplot \ref{fig:RFR_features}c shows two of the most important features plotted as a function of twist angle.

Each box in the decision-tree plot indicates a decision node with the feature used for dividing the feature space indicated at the top. In this example, feature \#188 of the input vectors (a particular intensity at the given location) is used and based on whether it is larger than 0.05 or not, the systems are divided into two groups. Analogous procedure continues at each node with decreasing mse between the predicted and the actual target value. The whole forest consists of many such trees, but with different features at each node. The decision tree structure allows to intuitively track how the model learns the twist angle from the input Raman spectra.

 The overall five most important features all correspond to intensities in close neighborhood of the G-band location. The error bars are given by std of importance across the forest. Among the five most important features, feature \#188 is the most informative, while the last two features \#168 and \#167 contain the least information. Mapping the feature index onto the wavenumber space shows that the most important feature corresponds to the intensity profile at 1672 cm$^{-1}$, only 5 cm$^{-1}$ up-shifted from the location of the G-band in pristine BLG. However, the intensity directly at the G-bands location for $\theta=0^{\circ}$, feature \#168, is not significant.

 Plotting the most important feature \#188 and feature \#168 as a function of the twist angle reveals some interesting patterns. These patterns were already partly visible in Fig.~\ref{fig:Spectra}, but their importance for MLRs couldn't be anticipated by visual inspection. Interestingly, feature \#188 shows peaks in its $\theta$-dependence at about 6$^{\circ}$ and 9$^{\circ}$ and vanishing intensity for $\theta>13^{\circ}$, while the points for angles below 5$^{\circ}$ form a cloud. In contrast, feature \#168 corresponds to higher intensity for $\theta>18^{\circ}$, a cloud of points for $\theta<5^{\circ}$ and comparably small intensity in between. The point clouds for small angles make it intuitively clear why small angles might be more difficult to identify from Raman spectra than their higher-$\theta$ counterparts.
 
 The most important features above can be used to reduce the dimensionality of the feature space. For example, RFR trained using only feature \#188 results in an average $R^2$ in ten-fold CV of 0.84 $\pm$ 0.06 and RMSE 3.29 $\pm$ 0.72. Including the first three most important features improves model performance to 0.93 $\pm$ 0.06 with RMSE 2.13 $\pm$ 0.88. Thus, using a very limited number of key features still results in a fairly good performance, but to achieve higher angle resolution as in table \ref{tab:scores} more spectral features have to be included.

\subsection{\label{subsec:Eresults}Discussion}
%%%--------------------
\begin{figure}[ht!]
\includegraphics[width=0.5\textwidth,keepaspectratio]
{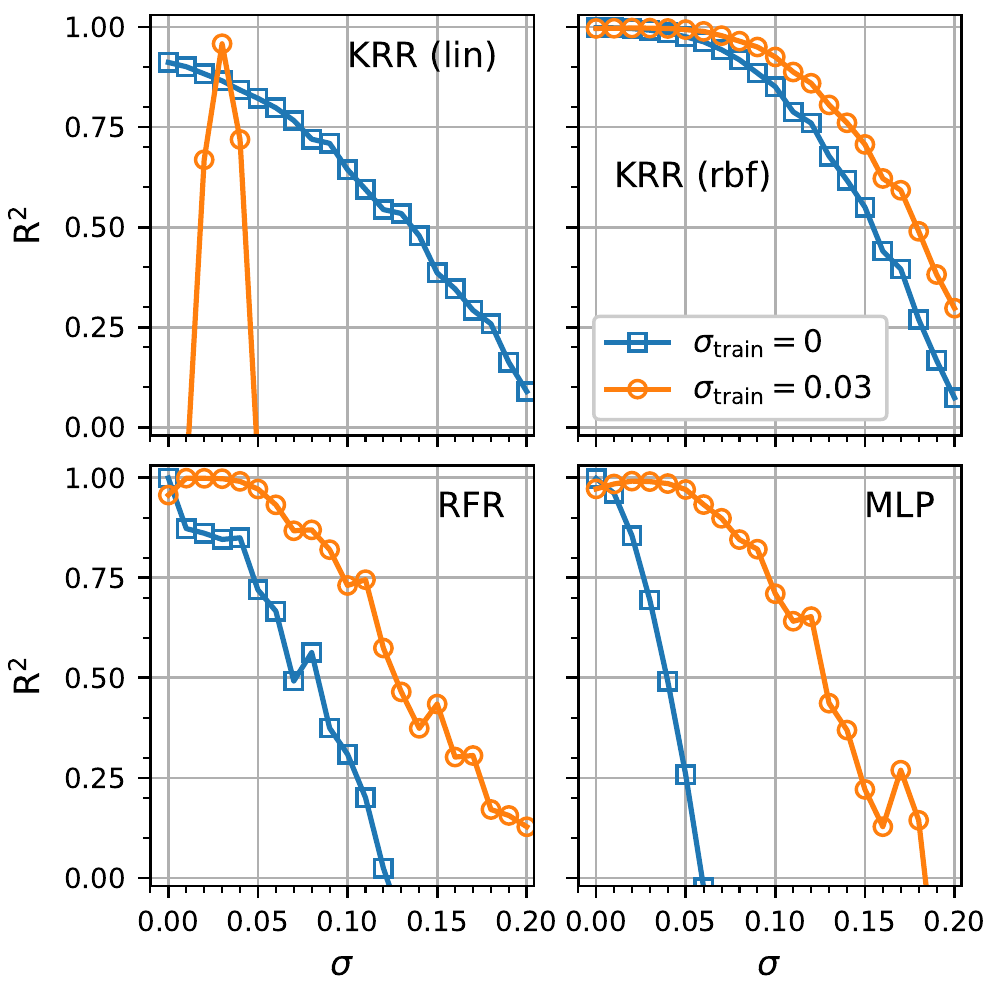}
\caption{\label{fig:noise_effect}
MLR performance-score $R^2$ as a function of $\sigma$. Gaussian noise with std $\sigma$ was added to the test data set. Comparison of two cases: MLRs trained on noiseless training data set ($\sigma_{\mathrm{train}}=0$, squares) \textit{vs.} training set augmented with small noise ($\sigma_{\mathrm{train}}=0.03$, circles).}
\end{figure}
%%%--------------------
The MLRs were trained on theoretical Raman spectra obtained computationally using a semi-empirical level of theory. Several aspects have to be discussed when it comes to the transferability of these models to experimental Raman spectra including the effect of noise in spectral data, the possibility of training ML models on associated phonon frequencies instead of intensities, and, most importantly, theory validation. 

\subsubsection{Effect of noise}
First, we tested the effect of noise on the MLRs' performance. Noise in experimental Raman spectra originates from multiple factors, mainly due to sample preparation and various aspects of the apparatus \cite{Barton2018}. As a result, experimental spectra collected under slightly different conditions might differ in their intensity profiles even for the same material. To simulate noise, Gaussian random noise with std $\sigma$ was added to the spectral intensity of the test data set for each point on the wavenumber grid. Figure~\ref{fig:noise_effect} shows the test $R^2$ coefficient resulting from applying the MLRs trained on noiseless data to noisy test data (squares). The values for $\sigma=0$ are identical to those reported in table \ref{tab:scores}. The performance score decreases with increasing $\sigma$ for all models, but at a different rate for each model. Interestingly, the KRR, both with linear and rbf kernels, is relatively stable against noise. The $R^2$ coefficient remains above 0.6 and even above 0.8 for the linear and rbf kernels, respectively in case of moderate noise with $\sigma=0.1$. In contrast, RFR's score quickly falls below 0.5 even for $\sigma=0.09$. For MLP, the model performance deteriorates even more dramatically with $R^2$ plunging below zero before $\sigma$ reached 0.07. Against the earlier expectation based on the learning-curve inspection, MLP turns out to be the least flexible model and the most strongly affected by overfitting. This is somewhat intuitive given that our database is relatively small (by ML standards) combined with the highly non-linear nature of MLP. Typical data sets that are better suitable for learning with ANNs contain at least a few thousand samples \cite{Carbone2019}. 

Next, artificial noise can be introduced directly during training in an attempt to make the models more robust. The resulting $R^2$ scores for $\sigma_{\mathrm{train}}=0.03$ as a function of noise in the test data are labeled by circles in Fig.~\ref{fig:noise_effect}. Clearly, the introduction of noise in the training set has a positive effect on all MLRs except the linear kernel KRR.
Especially for MLP and RFR there is remarkable improvement compared to $\sigma_{\mathrm{train}}=0$ with the performance score staying close to one up to $\sigma = 0.05$. Somewhat unsurprisingly, the linear kernel KRR has been "confused" by the presence of noise in the training set. The model performs very well on test data with $\sigma=\sigma_{\mathrm{train}}=0.03$, and acceptably so for $\sigma$ being close to the training noise (0.02 and 0.04). However, linear KRR fails completely as soon as the noise becomes considerably different from what the model has learned in training. This shows the rigidity of the model stemming from large weights being wrongly assigned to noise instead of actual spectral features. Ultimately, KRR with rbf kernel trained with $\sigma_{\mathrm{train}}=0.03$ is the most robust model. Its $R^2$ coefficient is larger than that of all other models particularly when $\sigma$ deviates significantly from $\sigma_{\mathrm{train}}$. 

\subsubsection{Using phonon frequencies}
\begin{figure}[ht!]
\includegraphics[width=0.5\textwidth,keepaspectratio]
{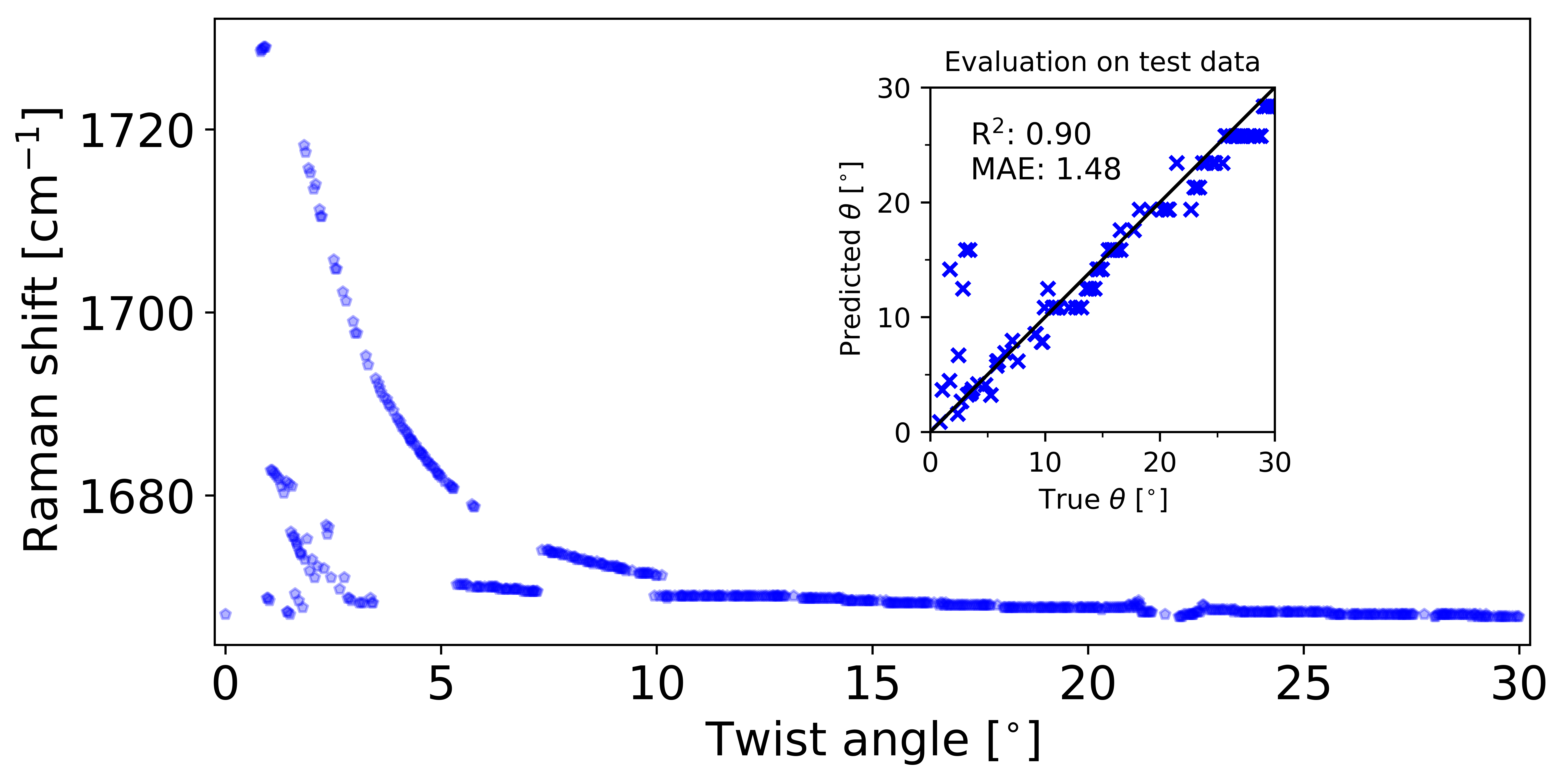}
\caption{\label{fig:Gbandplot}
Position of the most intense peak (G-band) as a function of twist angle $\theta$ (main plot). The inset shows the predicted \textit{vs.} true $\theta$ for KRR (rbf) model trained on the most intense peak's position evaluated on the holdout test set.
}
\end{figure}

Experimentalists typically process Raman spectra via peak fitting procedures in order to extract the frequencies of associated phonon modes. This approach was mimicked here by extracting the position of the most intense band as a function of twist angle, shown in Fig.~\ref{fig:Gbandplot}. This is essentially a way of informed dimensionality-reduction of the feature space. The $\omega(\theta)$ curve in Fig.~\ref{fig:Gbandplot} could almost be fitted to an exponential function if it wasn't for the outlier  points for $\theta<8^{\circ}$ that fall considerably off of the exponential trend. Using just the peak positions as features, a KRR (rbf) model was trained. The inset of Fig. \ref{fig:Gbandplot} shows the prediction of the resulting model \textit{vs.} true twist angle in the hold-out test set. In case of a perfect prediction, all points would lie exactly on the diagonal. There are some deviations, as one can see. Particularly, some small angles below $5^{\circ}$ are greatly overestimated, while larger angles $>10^{\circ}$ are mostly underestimated. This is intuitively clear from some $\omega(\theta)$ values being very similar for small as well as for larger twist angles. Nevertheless, even with this drastic dimensionality reduction to just one feature, the model's performance score is still relatively high (average $R^2$ in ten-fold CV is 0.87 $\pm$ 0.08). This highlights the importance of the "G-band" position variation for learning of the underlying twist angle.

%%%--------------------
%%%--------------------

\subsubsection{Focusing on the low-frequency spectral range}
Finally, one can test the possibility of analogously training MLRs using only spectral information in the LF region. As demonstrated earlier, relative to the strong G-band, the Raman intensities of lower-frequency phonon modes are essentially zero and thus carry no information for MLRs as demonstrated by PCA. Nonetheless, there are Raman active modes in the LF region that can now be measured with increasing accuracy. These modes (such as the shear and breathing modes) interrogate the interlayer interactions directly and are expected to probe the relative stackings between layers \cite{Liang2017}. Thus, alternative Raman spectra were constructed using only modes with frequencies below 200 cm$^{-1}$ and normalizing the intensities to the strongest peak within that region excluding the G-band information completely. Figure S5 shows the resulting spectra as a function of $\theta$. For $\theta<5^{\circ}$ there are some relatively strong intensity features in the ultra LF range that were observed before (see Fig. S1), but also additional strong peaks become clearly visible for larger angles starting from 80 cm$^{-1}$ as a result of the new normalization. Training KRR (rbf) and RFR on these spectra results in average (ten-fold CV) $R^2$ of 0.86 $\pm$ 0.03 and 0.97 $\pm$ 0.02, respectively. Interestingly, KRR does not perform as well as on the spectra data set containing only the G-band, while RFR has similar model performance as before. This might be due to the LF spectral features being strongly discontinuous in $\theta$, which RFR can handle better by dividing the feature space into "boxes" with nonlinear boundaries \cite{James2013}. Analogously to Fig. \ref{fig:RFR_features}, important features for RFR trained on the LF spectra database are demonstrated in Fig. S6. In this case, the most important feature corresponds to the intensity profile at 84 cm$^{-1}$. Although the important features carry less overall relative importance compared to those in Fig.~\ref{fig:RFR_features}, RFR is able to determine $\theta$ based on the collective intensity variation around 80 cm$^{-1}$ that together accounts for about 15\% of importance and has complementary $\theta$-dependence, see Fig. S6c. However, testing the effect of noise on models trained on the LF data set reveals significant overfitting on the part of RFR, see Fig. S7 [link to SM]. While RFR's performance score here is similar to that in Fig. \ref{fig:noise_effect} for $\sigma=0$, it plummets below zero as soon as any amount of noise is added to the test set. KRR (rbf), on the other hand, performs stably in the whole $\sigma$ range. In fact, KRR (rbf) trained on the LF data set turns out to be more robust to noise than its G-band-data-set counterpart, although the absolute performance scores of the former are lower than those of the latter for up to $\sigma=0.13$. For larger $\sigma$ values, KRR trained on the LF data set performs comparably or better than the same model trained on the G-band data set.

\subsubsection{Links to experiments}
The above discussions provide useful insights into the robustness and flexibility of the trained MLRs. KRR with rbf kernel turned out to be the most suitable approach to solve the present inverse problem using spectral information as input features. Additionally, reducing the dimensionality of the feature space to only the position of the most intense band can still provide reasonable, if more coarse-grain, information on the twist angle. However, the key issue in using the trained models in practice is theory validation. No matter how well the models perform on synthetic data, they can hardly be transferred to analysis of experimental spectra if the training data does not capture the experimental reality. Indeed, as briefly mentioned in the introduction, experimental Raman spectra of tBLG are quite complicated. Below, the Raman spectra calculated here are compared with experimental ones, focusing on the most prominent features, the G and the 2D bands. 

First, we start by noting the aspects that the present level of theory seems to capture. Experimentally, the G-band's position is said to remain practically constant with $\theta$. For this reason, the corresponding phonon frequency was set as a constant to the experimental value in previous theoretical studies \cite{Coh2013,Popov2018}. However, on a smaller scale, one can observe measurable, non-monotonic variations of a few wavenumbers in the G peak's position with twist angle (\textit{e.g.}, see Fig. 2 of Ref.~\cite{Kim2012}). These variations are well within experimental resolution. The present theoretical description highlights similar variations in figures \ref{fig:Spectra} and \ref{fig:Gbandplot}, assuming the most intense peak in each calculated spectrum is associated with the G-band, even though the absolute phonon frequency of the G-band was significantly overestimated. Moreover, the emergence of additional side bands close to the G-band for some $\theta$ \cite{Kim2012} could also be observed in the calculated spectra. Similarly, note that our phonon database contains even more additional modes in the range where experimental R bands are observed \cite{Jorio2013}. However, in contrast to experiment, their calculated Raman intensity is very small compared to that of the G-band so that the associated peaks are not visible in Fig. S1.

There are several experimental aspects that are not included in the present theoretical description. The Raman active phonons in our database with frequencies above the G-band frequency should not be identified as the experimental R' bands with one-to-one correspondence~\cite{Campos-Delgado2013}. This is because the REBO potential is known to fail at describing the Kohn anomalies in graphite and graphene, \textit{i.e.} the linear behavior of phonon bands close to $\Gamma$ and $K$ points of the Brillouin zone \cite{Maultzsch2004}. The linear behavior is attributed to electron-phonon coupling \cite{Maultzsch2004}, a phenomenon that is not captured within the present formalism. Consequently, the frequency of the experimental R' mode above the G-band that is attributed to the Brillouin zone folding of this higher frequency phonon on the zone center, where it becomes Raman active \cite{Campos-Delgado2013}, is not correctly predicted here.

In addition to the aforementioned difficulties in describing the phonon dispersion of the high-frequency branches \cite{Maultzsch2004,Lamparski2020}, the use of the REBO+KC potential results in a somewhat overestimated G-band frequency compared to experiment \cite{Kim2012}. Interestingly, for single-layer graphene, using REBO alone results in a somewhat down-shifted G-band frequency relative to experiment \cite{Zou2016,Rowe2018}. It seems the addition of the KC potential leads to a slight overall stiffening of the interatomic force-constants. In contrast, the REBO+KC potential apparently underestimates the interlayer force-constants slightly \cite{Lamparski2020}: The shear mode of AB-stacked BLG ($\theta=0^\circ$) is found at 19.7 cm$^{-1}$ (\textit{vs.} 32 cm$^{-1}$ in experiment \cite{Tan2012}), and the layer-breathing mode for $\theta=12^\circ$ is found at about 79 cm$^{-1}$ (\textit{vs.}  about 95 $^{-1}$ in experiment \cite{He2013}). We'd like to note here that the construction of accurate force-field potentials is the subject of intensive ongoing research efforts. Recently, ML was used to develop an accurate potential for graphene that has been reported to outperform other commonly used potentials such as REBO \cite{Brenner2002}, AIREBO \cite{Stuart2000}, LCBOP \cite{Los2003}, ReaxFF \cite{Duin2001} etc. as measured by the accuracy of the computed forces and phonons compared to DFT \cite{Rowe2018}. It would be interesting to test the present analysis in application to databases constructed with these other potentials. However, the construction of additional databases is beyond the scope of the present proof of concept study.

Next, several experimental aspects of the Raman intensities are not included in the theoretical description provided by the semi-empirical polarizability model. 
In relation to the G-band, its experimental intensity variation with $\theta$ is not well described. The relative intensity information between varying twist angles has been removed during spectra construction and normalization, see Sec.~\ref{subsec:Cmethods}. The normalization can be adjusted to preserve this information (by normalizing relative to the system's size). In this case, spectra belonging to structures with $\theta<5^{\circ}$ dominate the depiction of all the spectra analogous to Fig.~\ref{fig:Spectra}, and the spectra of the remaining systems are essentially vanishing in comparison. This contradicts the experimental findings where a intensity-enhancement is observed for $10^{\circ}<= \theta<15^{\circ}$ depending on laser energy \cite{Kim2012}. Even though experimental spectra of structures with angles below $5^{\circ}$ are rare, which might be a reason for an enhanced intensity for this case not being reported, the calculated $\theta$-dependent relative intensities should still be considered with caution. Moreover, the laser-energy dependence of the spectra is not included in the theoretical description used here and thus the experimentally observed resonance effects might not be described. Similarly, the width (FWHM) of the G-peak, that shows some non-monotonic variations with $\theta$ experimentally \cite{Mogera2020,Kim2012}, is not part of the present theoretical description. Generally, a peak's width is related to the lifetime of the associated phonon \cite{Coh2013,Popov2018} and requires additional considerations for theoretical modeling beyond the presently used semi-empirical model.

Another part of experimental Raman spectra of tBLG entirely missing in the present description is related to high-order peaks such as the 2D peak together with all its spectral properties position, laser-energy dependent intensity and width. These features of the 2D band exhibit even stronger variations with $\theta$ than the G-band \cite{Kim2012}. The 2D band stems from a two-phonon scattering process and its theoretical description therefore requires the inclusion of second-order Feynman diagrams \cite{Coh2013}. Thus, regardless of the employed force-field potentials, even if the use of DFT was feasible for tBLG, the entirety of information on the 2D peak would be missing in a first-order Raman scattering simulation.

Given the many experimental aspects missing from the present theoretical description, the trained MLRs cannot be reliably applied for analysis of experimental Raman spectra as is. Thus, it is necessary to employ more accurate methods for generation of the computational Raman spectra database. These results of these methods should then also be carefully compared with experimental observations. However, using more accurate theoretical models might prove difficult in practice, because of the rising computational expense that accompanies the complexity of the methods.

Similar problems of theory validation have been encountered in the simulation and machine learning of theoretical x-ray absorption near-edge structure (XANES) spectra for classification of local chemical environments \cite{Carbone2019}. It has been suggested that augmentation of the computational spectral database with available experimental data could allow for training of a more robust ML model. Training such a model would rely on ML techniques that are able to deal with a hybrid database, such as transfer learning \cite{Carbone2019}. Such an approach could potentially be useful for the present study of Raman spectra, as well. 

\section{\label{sec:conc}Conclusion}
We presented a computational framework for identifying the twist angle of tBLG systems from their Raman spectra corresponding to an inverse-problem solution. In addition to the construction of a structural and spectral database, the key goal of the present study was to establish the relationship between Raman intensities and the underlying twist angle of a graphene bilayer structure using machine learning regressors. All non-linear MLRs achieved high $R^{2}$ scores of 0.98 $\pm$ 0.04 and an average RMSE of 0.70$^{\circ}$ $\pm$ 0.93$^{\circ}$ and thus provided a continuous model of the twist angle with at least 2-degree resolution. Moreover, the feature-importance analysis unraveled that the intensity in the vicinity of the G-band of untwisted bilayer graphene is the most important feature for the trained ML models. In contrast, removing the intensities in the low-frequency regime did not affect model performance. Furthermore, the relative model's robustness upon the introduction of noise to the test data was tested. KRR with rbf kernel turned out to be the best performing model for this case. 

Despite the good performance and robustness of the MLRs trained on computational spectra, they cannot be transferred to experimental spectra as is. The relatively low level of theory that was used for database construction does not allow for description of key experimental features of tBLG systems. Nevertheless, this study provides a proof of concept showing that Raman spectra of the general shape similar to the calculated ones can be processed via MLRs in principle. Further improvements including higher-level theoretical descriptions and potentially the construction of a hybrid database containing both computational and experimental data are necessary to unambiguously obtain the twist angle of tBLG from experimental Raman spectra. Once accurate ML models that correctly capture the experimentally relevant features are available, the final framework would work as an executable black box. This black box would take in a measured Raman spectrum as input and return the underlying twist angle of the sample as identified by the ML model.

\section{\label{sec:thanks} Acknowledgments}
N.S. and M.L. contributed equally to this work. M.L performed computations to construct the structural and spectral database. N.S. devised and carried out the ML based analysis and prepared the manuscript. All authors discussed the results and contributed to the final manuscript. N.S. was supported by NSF Grant EFRI 2-DARE (EFRI-1542707).  
This work is also supported by the NY State Empire State Development's Division of Science, Technology and Innovation (NYSTAR) through Focus Center-NY–RPI Contract C150117.

\bibliography{references}
\end{document}